\documentclass[aps,preprint,showpacs,showkeys,nofootinbib,floatfix,superscriptaddress]{revtex4}
\usepackage{epsfig}
\usepackage{graphicx}
\usepackage{multirow}
\usepackage{tablefootnote}
\begin{document}

\title{\textbf{The $D^*$ $\Xi N$ bound state in strange three-body systems}}
\author{H.~Garcilazo} 
\email{humberto@esfm.ipn.mx} 
\affiliation{Escuela Superior de F\' \i sica y Matem\'aticas, \\ 
Instituto Polit\'ecnico Nacional, Edificio 9, 
07738 M\'exico D.F., Mexico} 

\author{A.~Valcarce} 
\email{valcarce@usal.es} 
\affiliation{Departamento de F\'\i sica Fundamental and IUFFyM,\\ 
Universidad de Salamanca, E-37008 Salamanca, Spain}

\date{\today} 

\begin{abstract}
The recent update of the strangeness $-2$ ESC08c Nijmegen potential 
incorporating the NAGARA and KISO events predicts 
a $\Xi N$ bound state, $D^*$, in the $^3S_1 (I=1)$
channel. We study if the existence of this two-body bound state could give rise
to stable three-body systems. For this purpose we solve the bound state 
problem of three-body systems where the $\Xi N$ state is merged
with $N$'s, $\Lambda$'s, $\Sigma's$ or $\Xi$'s, making use of the most recent updates 
of the two-body ESC08c Nijmegen potentials. We found that there appear 
stable states in the $\Xi NN$ and $\Xi \Xi N$ systems, the $\Xi \Lambda N$ and $\Xi \Sigma N$ 
systems being unbound.
\end{abstract}

\maketitle 

\section{Introduction}
The hyperon-nucleon ($YN$) and hyperon-hyperon ($YY$) interactions 
are not only of interest by themselves but they constitute also the
input for microscopic calculations of few- and many-body
systems involving strangeness, such as exotic 
neutron star matter~\cite{Dem10,Ant13,Wei12,Lon14,Mas15} or 
hypernuclei~\cite{Yam10,Hiy08,Yam01}.
It has been recently reported the so-called KISO event, the first 
clear evidence of a deeply bound state of $\Xi^{-} - ^{14}$N~\cite{Naa15}.
Although microscopic calculations are impossible in this case and, 
consequently, their interpretation will be always afflicted by large 
uncertainties, the ESC08c Nijmegen potential has been recently updated
to give account for the most recent experimental information of the strangeness
$-2$ sector, the KISO~\cite{Naa15} and the NAGARA~\cite{Tak01} events, concluding
the existence of a bound state, $D^*$, in the $^3S_1 (I=1)$ 
$\Xi N$ channel with a binding energy of 1.56 MeV~\cite{Nag15,Rij16}.

In a recent series of papers~\cite{Gar15,Gar16} we have studied the
consequences of the existence of this $\Xi N$ bound state in
few-baryon systems with nucleons, specifically because for some 
quantum numbers such states could be stable, what can be easily tested
against future data. In Ref.~\cite{Gar15} we have analyzed the 
possible existence  of $\Xi NN$ bound states in isospin $3/2$ channels, motivated
by the decoupling from the lowest $\Lambda\Lambda N$ channel, 
due to isospin conservation, what would make a possible bound 
state stable. We found a $\Xi NN$
$J^P=\frac{1}{2}^+$ bound state with a binding
energy of about 2.5 MeV\footnote{This binding energy was recalculated in Ref.~\cite{Gar16}
including in addition to the $\Xi N$ isospin-spin $(i,j)=(1,1)$ channel also the $\Xi N$ $(1,0)$ 
channel, which is mainly repulsive, obtaining a bound state with a binding energy slightly smaller,
1.33 MeV below threshold.}. In Ref.~\cite{Gar16} we found a 
$\Xi NN$ deeply bound state with quantum numbers $(I)J^P=(\frac{1}{2})\frac{3}{2}^+$, 
lying $13.5$ MeV below the $\Xi d$ threshold, due to the coherent effect of the deuteron, a $NN$
bound state, and the $D^*$, a $\Xi N$ bound state.

In a similar manner as the existence of the deuteron, a $NN$ bound state, is responsible for
the existence of the triton, $NNN$, and the hypertriton, $\Lambda NN$, stable three-body bound states,
in this paper we study if the existence of the $D^*$ $\Xi N$ bound state
could give rise to other stable few-body systems when it
is merged with $N$'s, $\Lambda$'s, $\Sigma$'s or $\Xi$'s. 
The possible existence of stable few-body states containing a $\Xi N$ two-body
subsystem is suggested by the attractive character of 
the $\Lambda \Xi$, $\Sigma \Xi$ and $\Xi\Xi$ interactions for some partial 
waves~\cite{Bea12,Sto99,Mil06,Hai10,Hai15,Nae15,Rij13}. 
There are also preliminary studies of the
$\Xi\Xi N$ system~\cite{Bea09} indicating that lattice QCD
calculations of multibaryon systems are now within sight.
To carry out our objectives, we will study
the $\Xi NN$, $\Xi \Lambda N$, $\Xi \Sigma N$ and $\Xi \Xi N$ three-body systems.
We will make use of the most recent updates of the
ESC08c Nijmegen potentials in the strangeness $-1$, $-2$, $-3$
and $-4$ sector~\cite{Nag15,Nae15,Rij13} accounting for the
recent KISO~\cite{Naa15} and NAGARA~\cite{Tak01} events in the strangeness 
$-2$ sector.

Recent preliminary results from lattice QCD suggest an overall attractive $\Xi N$ interaction~\cite{Sas15}
what may be relevant for the first $\Xi$ hypernucleus reported in Ref.~\cite{Naa15}.
Besides the recent update of ESC08c Nijmegen model~\cite{Nag15,Rij16}, there are other models
predicting bound states in the $\Xi N$ system previously to the KISO event, as
the chiral constituent quark model of Ref.~\cite{Car12}. 
However, one should keep in mind that there are other models 
for the $\Xi N$ interaction, like the hybrid 
quark--model based analysis of Ref.~\cite{Fuj07}, the effective field 
theory approach of Ref.~\cite{Hai16}, or even some of the
earlier models of the Nijmegen group~\cite{Sto99} that do not present $\Xi N$
bound states and, in general, the interactions are weakly attractive or repulsive.
Thus, one does not expect that these models will give rise to 
$\Xi NN$ or $\Xi YN$ bound states. On the other hand, 
current $\Xi$ hypernuclei studies~\cite{Yam10,Hiy08,Yam01} 
have been performed by means of $\Xi N$ interactions derived from the 
Nijmegen models and thus our study complements such previous works 
for the simplest systems that could be studied exactly.

The paper is organized as follows. We will use Sec.~\ref{secII} for describing
all technical details to solve the three-body bound-state Faddeev equations. 
In Sec.~\ref{secIII} we will construct the two-body amplitudes needed for the
solution of the bound state three-body problem. Our results will be 
presented and discussed in Sec.~\ref{secIV}.
Finally, in Sec.~\ref{secV} we summarize our main conclusions.

\section{The three-body bound-state Faddeev equations}
\label{secII}

We will restrict ourselves to the configurations where all three 
particles are in S-wave states so that the Faddeev equations 
for the bound-state problem in the case of three baryons with total
isospin $I$ and total spin $J$ are,
\begin{eqnarray}
T_{i;IJ}^{i_ij_i}(p_iq_i) = &&\sum_{j\ne i}\sum_{i_jj_j}
h_{ij;IJ}^{i_ij_i;i_jj_j}\frac{1}{2}\int_0^\infty q_j^2dq_j
\int_{-1}^1d{\rm cos}\theta\, 
t_{i;i_ij_i}(p_i,p_i^\prime;E-q_i^2/2\nu_i) 
\nonumber \\ &&
\times\frac{1}{E-p_j^2/2\mu_j-q_j^2/2\nu_j}\;
T_{j;IJ}^{i_jj_j}(p_jq_j) \, , 
\label{eq1} 
\end{eqnarray}
where $t_{i;i_ij_i}$ stands for the two-body amplitudes
with isospin $i_i$ and spin $j_i$. $p_i$
is the momentum of the pair $jk$ (with $ijk$ an even permutation of
$123$) and $q_i$ the momentum of particle $i$ with respect to the pair
$jk$. $\mu_i$ and $\nu_i$ are the corresponding reduced masses,
\begin{eqnarray}
\mu_i &=& \frac{m_jm_k}{m_j+m_k} \, , \nonumber\\
\nu_i &=& \frac{m_i(m_j+m_k)}{m_i+m_j+m_k} \, ,
\label{eq3}
\end{eqnarray}
and the momenta $p_i^\prime$ and $p_j$ in Eq.~(\ref{eq1}) are given by,
\begin{eqnarray}
p_i^\prime &=& \sqrt{q_j^2+\frac{\mu_i^2}{m_k^2}q_i^2+2\frac{\mu_i}{m_k}
q_iq_j{\rm cos}\theta} \, , \nonumber \\
p_j &=& \sqrt{q_i^2+\frac{\mu_j^2}{m_k^2}q_j^2+2\frac{\mu_j}{m_k}
q_iq_j{\rm cos}\theta} \, .
\label{eq5}
\end{eqnarray}
$h_{ij;IJ}^{i_ij_i;i_jj_j}$ are the spin--isospin coefficients,
\begin{eqnarray}
h_{ij;IJ}^{i_ij_i;i_jj_j}= &&
(-)^{i_j+\tau_j-I}\sqrt{(2i_i+1)(2i_j+1)}
W(\tau_j\tau_kI\tau_i;i_ii_j)
\nonumber \\ && \times
(-)^{j_j+\sigma_j-J}\sqrt{(2j_i+1)(2j_j+1)}
W(\sigma_j\sigma_kJ\sigma_i;j_ij_j) \, , 
\label{eq6}
\end{eqnarray}
where $W$ is the Racah coefficient and $\tau_i$, $i_i$, and $I$ 
($\sigma_i$, $j_i$, and $J$) are the isospins (spins) of particle $i$,
of the pair $jk$, and of the three--body system.

Since the variable $p_i$ in Eq.~(\ref{eq1}) runs from 0 to $\infty$,
it is convenient to make the transformation
\begin{equation}
x_i=\frac{p_i-b}{p_i+b} \, ,
\label{eq7}
\end{equation}
where the new variable $x_i$ runs from $-1$ to $1$ and $b$ is a scale
parameter that has no effect on the solution. With this transformation
Eq.~(\ref{eq1}) takes the form,
\begin{eqnarray}
T_{i;IJ}^{i_ij_i}(x_iq_i) = &&\sum_{j\ne i}\sum_{i_jj_j}
h_{ij;IJ}^{i_ij_i;i_jj_j}\frac{1}{2}\int_0^\infty q_j^2dq_j
 \int_{-1}^1d{\rm cos}\theta\; 
t_{i;i_ij_i}(x_i,x_i^\prime;E-q_i^2/2\nu_i) 
\nonumber \\ &&
\times\frac{1}{E-p_j^2/2\mu_j-q_j^2/2\nu_j}\;
T_{j;IJ}^{i_jj_j}(x_jq_j) \, . 
\label{eq8} 
\end{eqnarray}
Since in the amplitude $t_{i;i_ij_i}(x_i,x_i^\prime;e)$ the variables
$x_i$ and $x_i^\prime$ run from $-1$ to $1$, one can expand this amplitude
in terms of Legendre polynomials as,
\begin{equation}
t_{i;i_ij_i}(x_i,x_i^\prime;e)=\sum_{nr}P_n(x_i)\tau_{i;i_ij_i}^{nr}(e)P_r(x'_i) \, ,
\label{eq9}
\end{equation}
where the expansion coefficients are given by,
\begin{equation}
\tau_{i;i_ij_i}^{nr}(e)= \frac{2n+1}{2}\frac{2r+1}{2}\int_{-1}^1dx_i
\int_{-1}^1 dx_i^\prime\; P_n(x_i) 
t_{i;i_ij_i}(x_i,x_i^\prime;e)P_r(x_i^\prime) \, .
\label{eq10} 
\end{equation}
Applying expansion~(\ref{eq9}) in Eq.~(\ref{eq8}) one gets,
\begin{equation}
T_{i;IJ}^{i_ij_i}(x_iq_i) = \sum_n P_n(x_i) T_{i;IJ}^{ni_ij_i}(q_i) \, ,
\label{eq11}
\end{equation}
where $T_{i;IJ}^{ni_ij_i}(q_i)$ satisfies the one-dimensional integral equation,
\begin{equation}
T_{i;IJ}^{ni_ij_i}(q_i) = \sum_{j\ne i}\sum_{mi_jj_j}
\int_0^\infty dq_j A_{ij;IJ}^{ni_ij_i;mi_jj_j}(q_i,q_j;E)\;
T_{j;IJ}^{mi_jj_j}(q_j) \, , 
\label{eq12}
\end{equation}
with
\begin{eqnarray}
A_{ij;IJ}^{ni_ij_i;mi_jj_j}(q_i,q_j;E)= &&
h_{ij;IJ}^{i_ij_i;i_jj_j}\sum_r\tau_{i;i_ij_i}^{nr}(E-q_i^2/2\nu_i)
\frac{q_j^2}{2}
\nonumber \\ &&
\times\int_{-1}^1 d{\rm cos}\theta\;\frac{P_r(x_i^\prime)P_m(x_j)} 
{E-p_j^2/2\mu_j-q_j^2/2\nu_j} \, .
\label{eq13} 
\end{eqnarray}

The three amplitudes $T_{1;IJ}^{ri_1j_1}(q_1)$, $T_{2;IJ}^{mi_2j_2}(q_2)$,
and $T_{3;IJ}^{ni_3j_3}(q_3)$ in Eq.~(\ref{eq12}) are coupled together.
The number of coupled equations can be reduced, however, when two of
the particles are identical. The reduction procedure for the case where
one has two identical fermions has been described before~\cite{Afn74,Gar90}
and will not be repeated here. With the assumption that 
particles 2 and 3 are identical and
particle 1 is the different one, only the amplitudes 
$T_{1;IJ}^{ri_1j_1}(q_1)$ and $T_{2;IJ}^{mi_2j_2}(q_2)$ are independent
from each other and they satisfy the coupled integral equations,
\begin{equation}
T_{1;IJ}^{ri_1j_1}(q_1)  =  2\sum_{mi_2j_2}
\int_0^\infty dq_3 A_{13;IJ}^{ri_1j_1;mi_2j_2}(q_1,q_3;E)\;
T_{2;IJ}^{mi_2j_2}(q_3) \, ,
\label{eq14} 
\end{equation}
\begin{eqnarray}
T_{2;IJ}^{ni_2j_2}(q_2) = && \sum_{mi_3j_3}g
\int_0^\infty dq_3 A_{23;IJ}^{ni_2j_2;mi_3j_3}(q_2,q_3;E)\;
T_{2;IJ}^{mi_3j_3}(q_3) 
\nonumber \\ && +
\sum_{ri_1j_1}
\int_0^\infty dq_1 A_{31;IJ}^{ni_2j_2;ri_1j_1}(q_2,q_1;E)\;
T_{1;IJ}^{ri_1j_1}(q_1) \, , 
\label{eq15} 
\end{eqnarray}
with the identical--particle factor
\begin{equation}
g=(-)^{1+\sigma_1+\sigma3-j_2+\tau_1+\tau_3-i_2} \, ,
\label{eq16}
\end{equation}
where $\sigma_1$ ($\tau_1$) 
stand for the spin (isospin) of the different particle and
$\sigma_3$ ($\tau_3$) for those of the identical ones.

Substitution of Eq.~(\ref{eq14}) into Eq.~(\ref{eq15}) yields an
equation with only the amplitude $T_2$,
\begin{equation}
T_{2;IJ}^{ni_2j_2}(q_2) = \sum_{mi_3j_3}
\int_0^\infty dq_3 K_{IJ}^{ni_2j_2;mi_3j_3}(q_2,q_3;E)\;
T_{2;IJ}^{mi_3j_3}(q_3) \, , 
\label{eq17}
\end{equation}
where
\begin{eqnarray}
K_{IJ}^{ni_2j_2;mi_3j_3}(q_2,q_3;E)= && g
A_{23;IJ}^{ni_2j_2;mi_3j_3}(q_2,q_3;E)
\nonumber \\ && +
2\sum_{ri_1j_1}
\int_0^\infty dq_1 A_{31;IJ}^{ni_2j_2;ri_1j_1}(q_2,q_1;E)
A_{13;IJ}^{ri_1j_1;mi_3j_3}(q_1,q_3;E) \, .
\label{eq18} 
\end{eqnarray}

\section{Two--body amplitudes}
\label{secIII}

We have constructed the two-body amplitudes for all subsystems entering the three-body problems
studied by solving the Lippmann--Schwinger
equation of each $(i,j)$ channel,
\begin{equation}
t^{ij}(p,p';e)= V^{ij}(p,p')+\int_0^\infty {p^{\prime\prime}}^2
dp^{\prime\prime} V^{ij}(p,p^{\prime\prime})
\frac{1}{e-{p^{\prime\prime}}^2/2\mu} t^{ij}(p^{\prime\prime},p';e) \, ,
\label{eq19} 
\end{equation}
where 
\begin{equation}
V^{ij}(p,p')=\frac{2}{\pi}\int_0^\infty r^2dr\; j_0(pr)V^{ij}(r)j_0(p'r) \, ,
\label{eq20} 
\end{equation}
and the two-body potentials consist of an attractive and a repulsive
Yukawa term, i.e.,
\begin{equation}
V^{ij}(r)=-A\frac{e^{-\mu_Ar}}{r}+B\frac{e^{-\mu_Br}}{r} \, .
\label{eq21} 
\end{equation}
The parameters of all $\Lambda N$, $\Sigma N$, $\Xi N$, $\Lambda \Xi$, $\Sigma \Xi$ and $\Xi \Xi$ 
channels were obtained by fitting the low-energy data of each channel as given 
in the most recent update of the strangeness $-1$ and $-2$~\cite{Nag15} and strangeness $-3$ and $-4$~\cite{Rij13} ESC08c Nijmegen 
potential. The low-energy data and the parameters of these models are given in Table~\ref{t1}. 
The $\Xi N$ $^1S_0$ $(I=0)$ potential was fitted to the $\Xi N$ phase shifts given in Fig. 14 
of Ref.~\cite{Nag15} without taking into account the inelasticity, i.e., assuming $\rho=0$
(this two-body channel does not contribute to the three-body bound states found
in this work). For the $\Sigma N$ system we only consider the $I=3/2$ channels, because the
$I=1/2$ channels would decay strongly to $\Lambda N$ states. Analogously,
for the $\Sigma \Xi$ system we only consider the $I=3/2$ channels, because the
$I=1/2$ channels would decay strongly to $\Lambda \Xi$ states. In the case 
of the $NN$ $(0,1)$ and $(1,0)$ channels we use
the Malfliet-Tjon models~\cite{Mal69} with the parameters given in Ref.~\cite{Gib90}.
\begin{table}[t]
\caption{Low-energy parameters of the most recent updates of the ESC08c Nijmegen interactions
for the $\Lambda N$~\cite{Nag15}, $\Sigma N$~\cite{Nag15}, $\Xi N$~\cite{Nag15}, $\Lambda \Xi$~\cite{Rij13}, $\Sigma \Xi$~\cite{Rij13} and
$\Xi \Xi$~\cite{Rij13} systems, and the parameters of the corresponding local potentials given
by Eq.~(\ref{eq21}).} 
\begin{ruledtabular} 
\begin{tabular}{ccccccccc} 
& $(i,j)$ & $a({\rm fm})$ & $r_0({\rm fm})$ & $A$(MeV fm) & 
$\mu_A({\rm fm}^{-1}$) 
& $B$(MeV fm) & $\mu_B({\rm fm}^{-1})$  & \\
\hline
\multirow{2}{*}{$\Lambda N$}& $(1/2,0)$ & $-2.62$  & $3.17$  &  $280$  & $2.00$  & $655$ & $3.55$ & \\ 
& $(1/2,1)$ & $-1.72$  & $3.50$  &  $170$  & $1.95$  & $670$ & $4.60$ & \\ \hline
\multirow{2}{*}{$\Sigma N$}& $(3/2,0)$ & $-3.91$  & $3.41$  &  $122$  & $1.47$  & $388$ & $3.55$ & \\ 
& $(3/2,1)$ & $0.61$  & $-2.35$  &  $329$  & $4.12$  & $124$ & $1.71$ & \\ \hline
\multirow{4}{*}{$\Xi N$}
& $(0,0)$\footnotemark[1] & $-$      & $-$      &  $120$  & $1.30$  & $510$ & $2.30$ & \\
& $(0,1)$ & $-5.357$ & $1.434$  &  $377$  & $2.68$  & $980$ & $6.61$ & \\ 
& $(1,0)$ & $0.579$  & $-2.521$ &  $290$  & $3.05$  & $155$ & $1.60$ & \\
& $(1,1)$ & $4.911$  & $0.527$  &  $568$  & $4.56$  & $425$ & $6.73$ & \\ \hline
\multirow{2}{*}{$\Lambda \Xi$} & $(1/2,0)$ & $-9.83$ & $2.38$  &  $370$  & $2.20$  & $970$ & $3.90$ & \\ 
& $(1/2,1)$ & $-12.9$ & $2.00$  &  $130$  & $1.90$  & $340$ & $4.50$ & \\ \hline
\multirow{2}{*}{$\Sigma \Xi$} & $(3/2,0)$ & $-2.80$ & $2.45$  &  $111$  & $2.00$  & $315$ & $4.73$ & \\ 
& $(3/2,1)$ & $-10.9$ & $1.92$  &  $147$  & $2.07$  & $790$ & $6.33$ & \\ \hline
\multirow{2}{*}{$\Xi \Xi$} & $(0,1)$ & $0.53$   & $1.63$  &  $210$  & $1.60$  & $560$ & $2.05$ & \\ 
& $(1,0)$ & $-7.25$  & $2.00$  &  $155$  & $1.75$  & $490$ & $5.60$ & \\
\end{tabular}
\footnotetext[1]{This channel is discussed on Sec.~\ref{secIII}.}
\end{ruledtabular}
\label{t1} 
\end{table}
\begin{figure*}[t]
\resizebox{8.cm}{12.cm}{\includegraphics{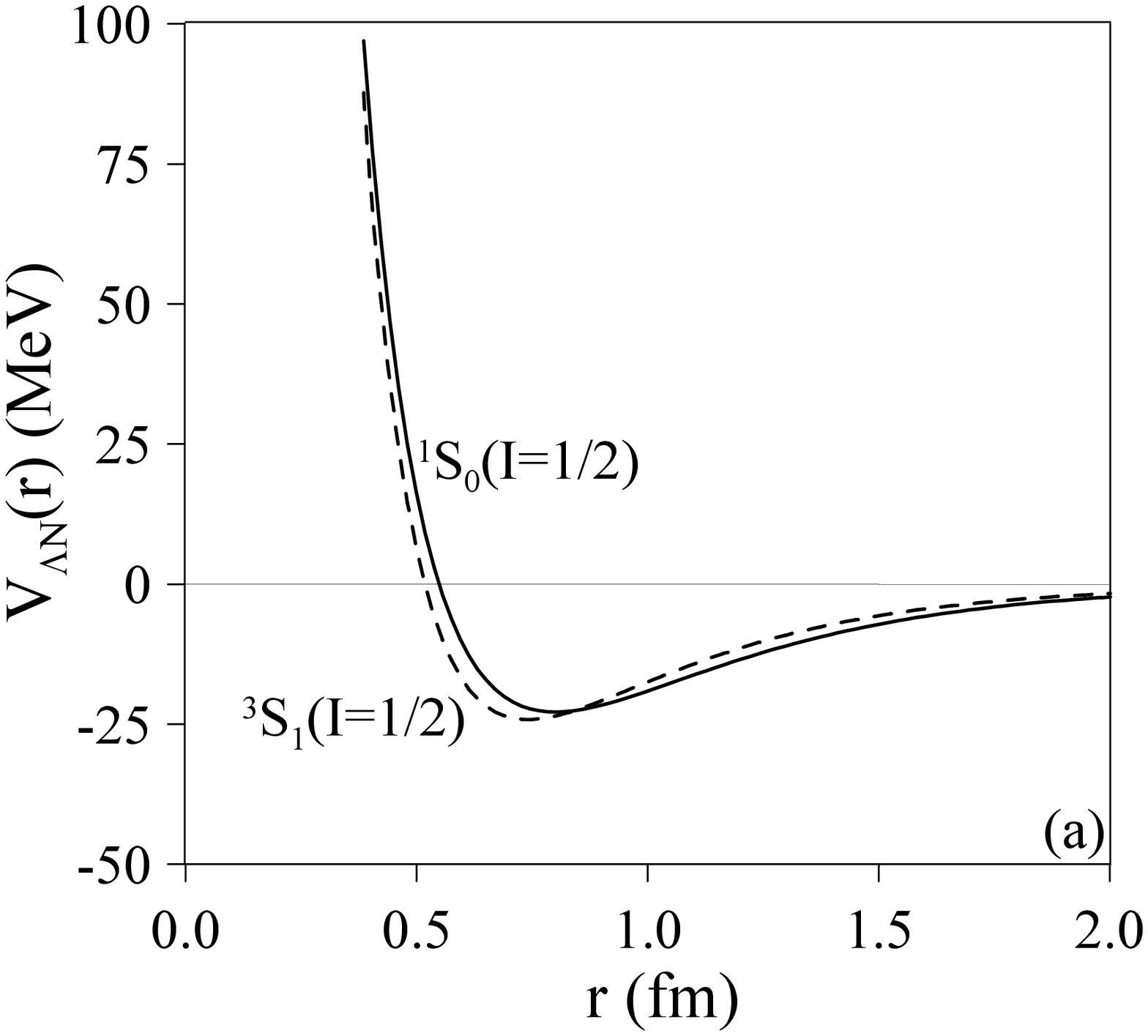}}
\resizebox{8.cm}{12.cm}{\includegraphics{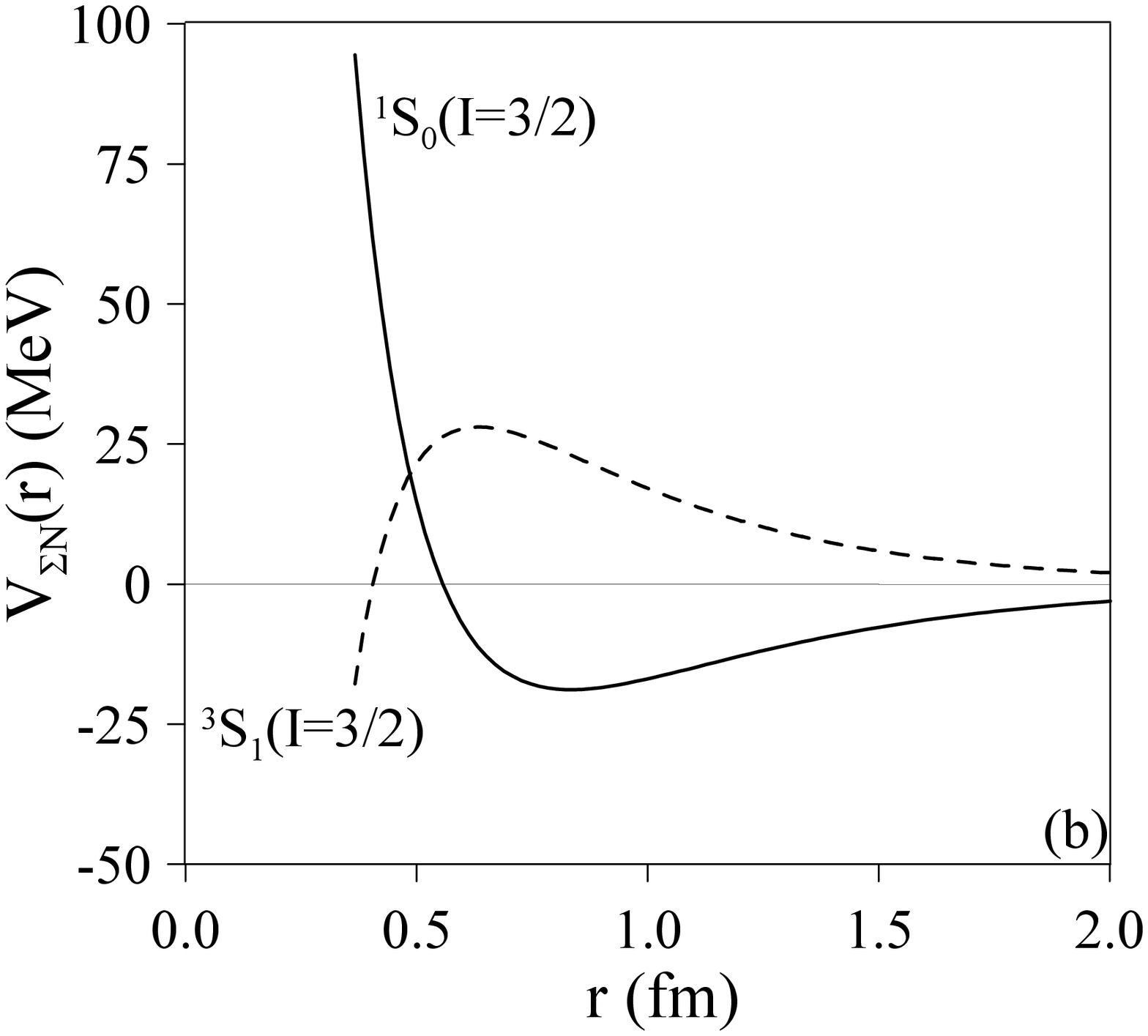}}\vspace*{-5.0cm}
\resizebox{8.cm}{12.cm}{\includegraphics{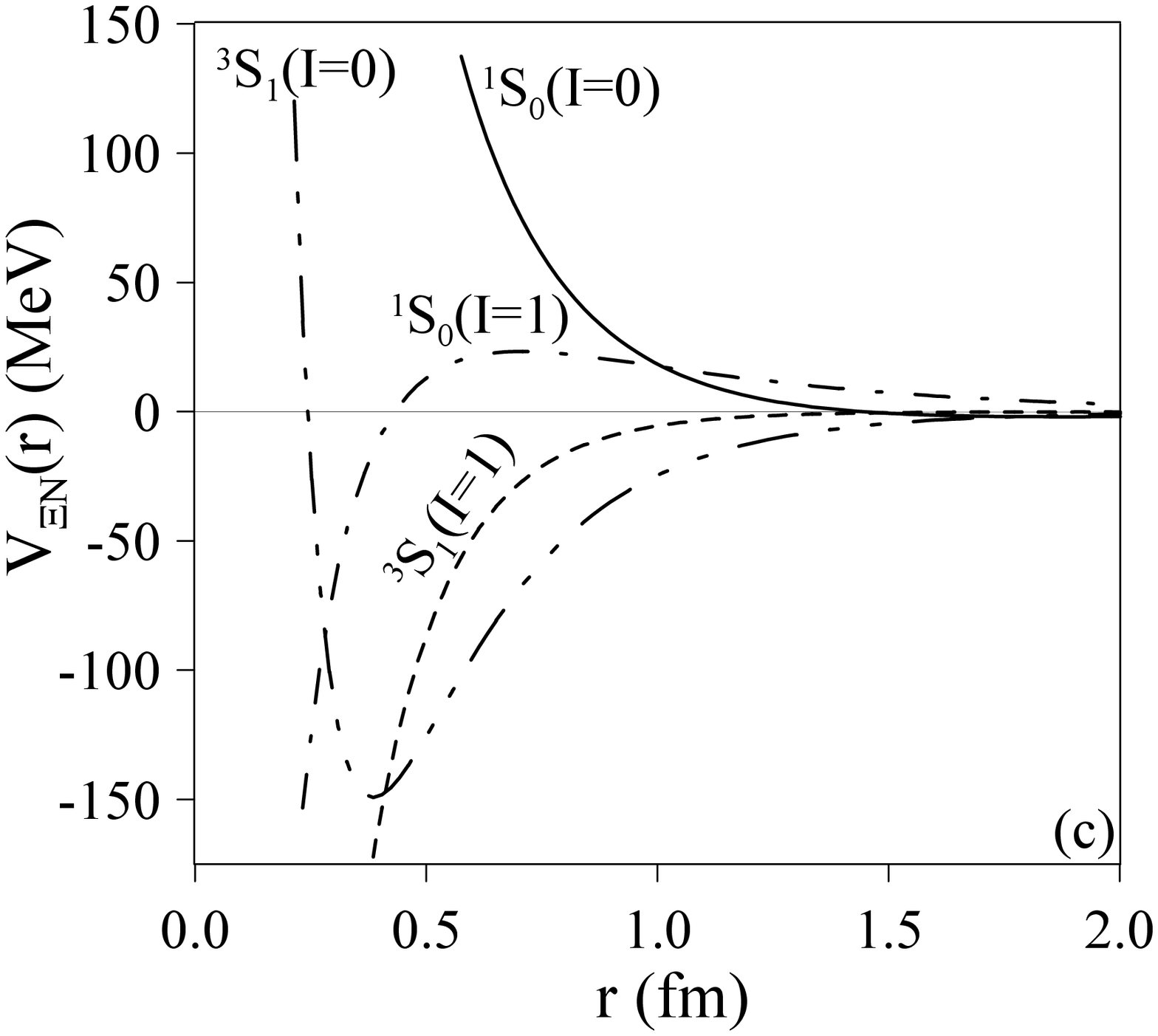}}
\resizebox{8.cm}{12.cm}{\includegraphics{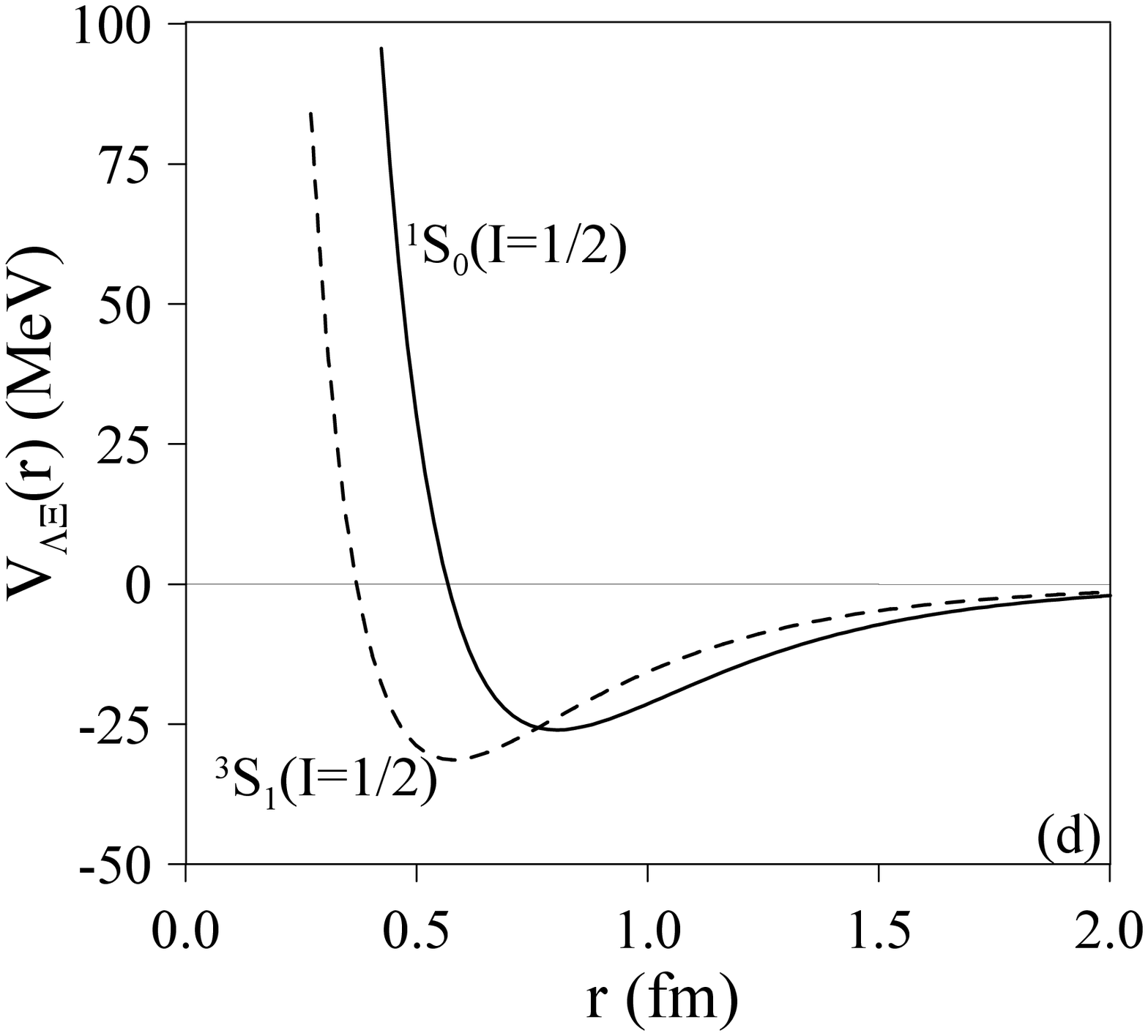}}\vspace*{-5.0cm}
\resizebox{8.cm}{12.cm}{\includegraphics{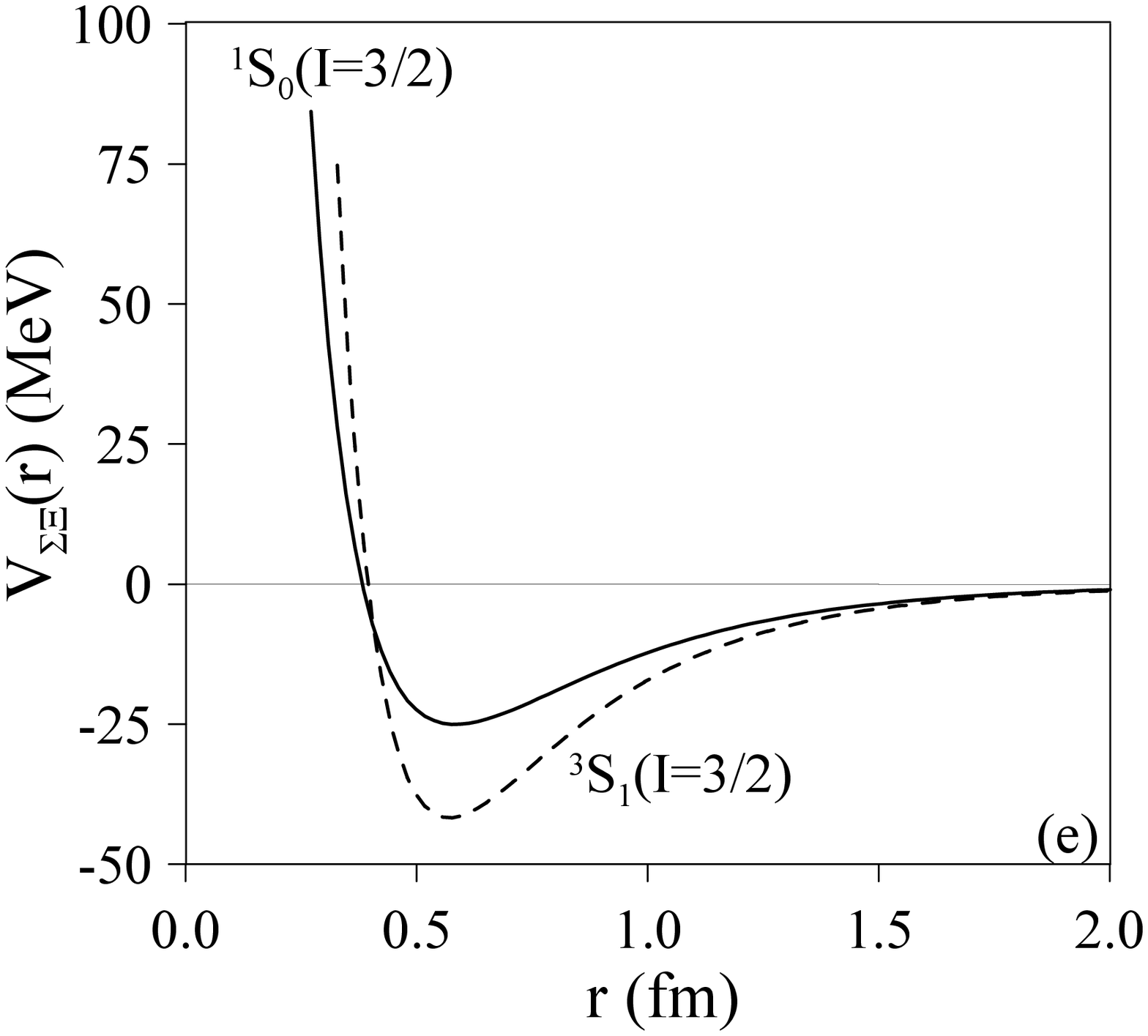}}
\resizebox{8.cm}{12.cm}{\includegraphics{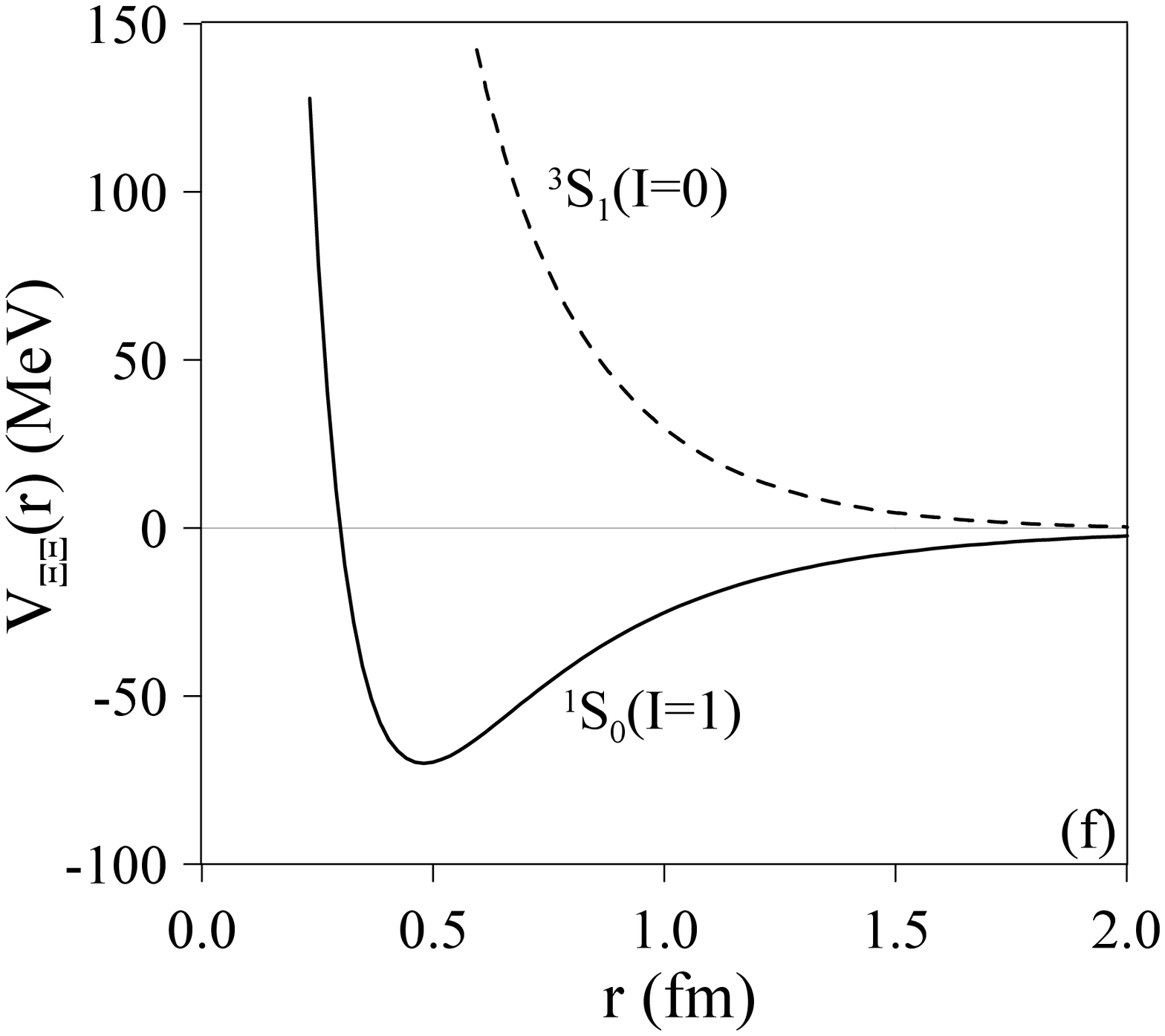}}
\vspace*{-5.0cm}
\caption{(a) $V_{\Lambda N}(r)$ potential as given by Eq.~(\ref{eq21}) with the
parameters of Table~\ref{t1}.
(b) Same as (a) for the $V_{\Sigma N}(r)$ potential.
(c) Same as (a) for the $V_{\Xi N}(r)$ potential.
(d) Same as (a) for the $V_{\Lambda \Xi}(r)$ potential.
(e) Same as (a) for the $V_{\Sigma \Xi}(r)$ potential.
(f) Same as (a) for the$V_{\Xi\Xi}(r)$ potential.}
\label{fig1}
\end{figure*}

The potentials obtained are shown in Fig.~\ref{fig1}. In Fig.~\ref{fig1}(a) we show the
$V_{\Lambda N}(r)$ potential that it is constrained by the existence of experimental data.
The interaction is attractive at intermediate range and strongly repulsive at short range,
but without having bound states. The same could be said about the $I=3/2$ $V_{\Sigma N}(r)$ 
potentials shown in Fig.~\ref{fig1}(b). The existence of $\Sigma^\pm p$ cross sections
tightly constrains the interaction. As can be seen the $^3S_1(I=3/2)$ potential is
strongly repulsive at intermediate range, what makes rather unlikely the existence
of three-body bound states containing this $\Sigma N$ channel.
In Fig.~\ref{fig1}(c) we show the $V_{\Xi N}(r)$ potential,
where one notes the attractive character of the $^3S_1(I=1)$ $\Xi N$ partial wave, 
giving rise to the $D^*$ bound state with a binding energy of 1.67 MeV. 
We also confirm how all the $J=1$ $\Xi N$ interactions are attractive~\cite{Rij13}.
The $V_{\Lambda \Xi}(r)$ potentials shown in Fig.~\ref{fig1}(d) are rather similar 
to the $V_{\Lambda N}(r)$ case, the intermediate range attraction not being enough
to generate two-body bound states. 
The $I=3/2$ $V_{\Sigma \Xi}(r)$ potentials are shown in Fig.~\ref{fig1}(e),
analogously to the $\Lambda \Xi$ case,  being attractive they do not present two-body bound states.
Regarding the $\Xi\Xi$ interaction, Fig.~\ref{fig1}(f), we
observe the attractive character of the $^1S_0(I=1)$ potential, that although having 
bound states in earlier versions of the ESC08c Nijmegen potential~\cite{Sto99}, in the most recent
update of the strangeness $-4$ sector it does not present a bound state~\cite{Rij13}.
The existence of bound states in the $\Xi\Xi$ system has been predicted by different 
calculations in the literature~\cite{Bea12,Mil06,Hai10}. What can be definitively stated 
that all models agree it is on the fairly important attractive character of this 
channel either with a bound state or not~\cite{Hai15}.

\section{Results and discussion}
\label{secIV}
\begin{table}[t]
\caption{Two-body $NN$, $YN$ and $YY$ isospin-spin $(i,j)$ channels that contribute to a 
given three-body state with total isospin $I$ and total spin $J$. The last column
indicates the corresponding threshold for each state, that would come given by $\sum_{i=1}^3 M_i - E$,
where $M_i$ are the masses of the baryons of each channel, $B_1$ stands for the binding energy of 
the deuteron and $B_2$ for the binding energy of the $D^*$ $\Xi N$ state.}
\resizebox{16.5cm}{!} {
\begin{tabular}{ccccccccc} \hline\hline
& $(I,J)$ & $\Lambda N$ & $\Xi N$ & $\Lambda \Xi$ & $\Sigma \Xi (\Sigma N)$ & $\Xi \Xi(NN)$ & $E$ \\
\hline
\multirow{4}{*}{$\Xi NN$}
& $(1/2,1/2)$ & $-$ & (0,0),(0,1),(1,0),(1,1) &$-$ & $-$ & (0,1),(1,0) & $B_1$ \\
& $(1/2,3/2)$ & $-$ & (0,1),(1,1) & $-$  & $-$ & (0,1) &  $ B_1$ \\
& $(3/2,1/2)$ & $-$ & (1,0),(1,1) & $-$  & $-$ & (1,0) & $ B_2$ \\
& $(3/2,3/2)$ & $-$ & (1,1) & $-$ & $-$ & $-$ &   $ B_2$ \\ \hline
\multirow{4}{*}{$\Xi \Lambda N$}
& $(0,1/2)$ & (1/2,0),(1/2,1) & (0,0),(0,1) & (1/2,0),(1/2,1) & $-$ & $-$ & $ 0 $ \\
& $(0,3/2)$ & (1/2,1)  & (0,1) & (1/2,1) & $-$ & $-$ & $ 0 $ \\
& $(1,1/2)$ & (1/2,0),(1/2,1) & (1,0),(1,1) & (1/2,0),(1/2,1) & $-$ & $-$ & $ B_2$ \\
& $(1,3/2)$ & (1/2,1)  & (1,1) & (1/2,1) & $-$ & $-$ & $ B_2$ \\ \hline
\multirow{2}{*}{$\Xi \Sigma N$}
& $(2,1/2)$ & $-$ & (1,0),(1,1) & $-$ & (3/2,0),(3/2,1) &  $-$ & $ B_2$ \\
& $(2,3/2)$ & $-$ & (1,1) & $-$ & (3/2,1) & $-$ & $ B_2$ \\ \hline
\multirow{4}{*}{$\Xi \Xi N$} 
& $(1/2,1/2)$ & $-$ & (0,0),(0,1),(1,0),(1,1) & $-$ & $-$ & (0,1),(1,0) & $ B_2 $ \\
& $(1/2,3/2)$ & $-$ & (0,1),(1,1) & $-$ & $-$ & (0,1) &  $ B_2$  \\
& $(3/2,1/2)$ & $-$ & (1,0),(1,1) & $-$ & $-$ & (1,0) & $ B_2$ \\
& $(3/2,3/2)$ & $-$ & (1,1) & $-$ & $-$ & $-$ & $ B_2$ \\ \hline\hline
\end{tabular}
}
\label{t2} 
\end{table}

We show in Table~\ref{t2} the channels of the different two-body subsystems contributing
to each $(I,J)$ three-body state. For the $\Xi\Sigma N$ system we only consider
the $I=2$ channels, because the $I=0$ and $1$ would decay strongly to $\Xi\Lambda N$ states.
The three-body problem is solved by means
of the ESC08c Nijmegen interactions described in Sec.~\ref{secIII} and given in Table~\ref{t1}. The binding energies are
measured with respect to the lowest threshold, indicated in Table~\ref{t2} 
for each particular state. 

\begin{figure*}[t]
\resizebox{8.cm}{12.cm}{\includegraphics{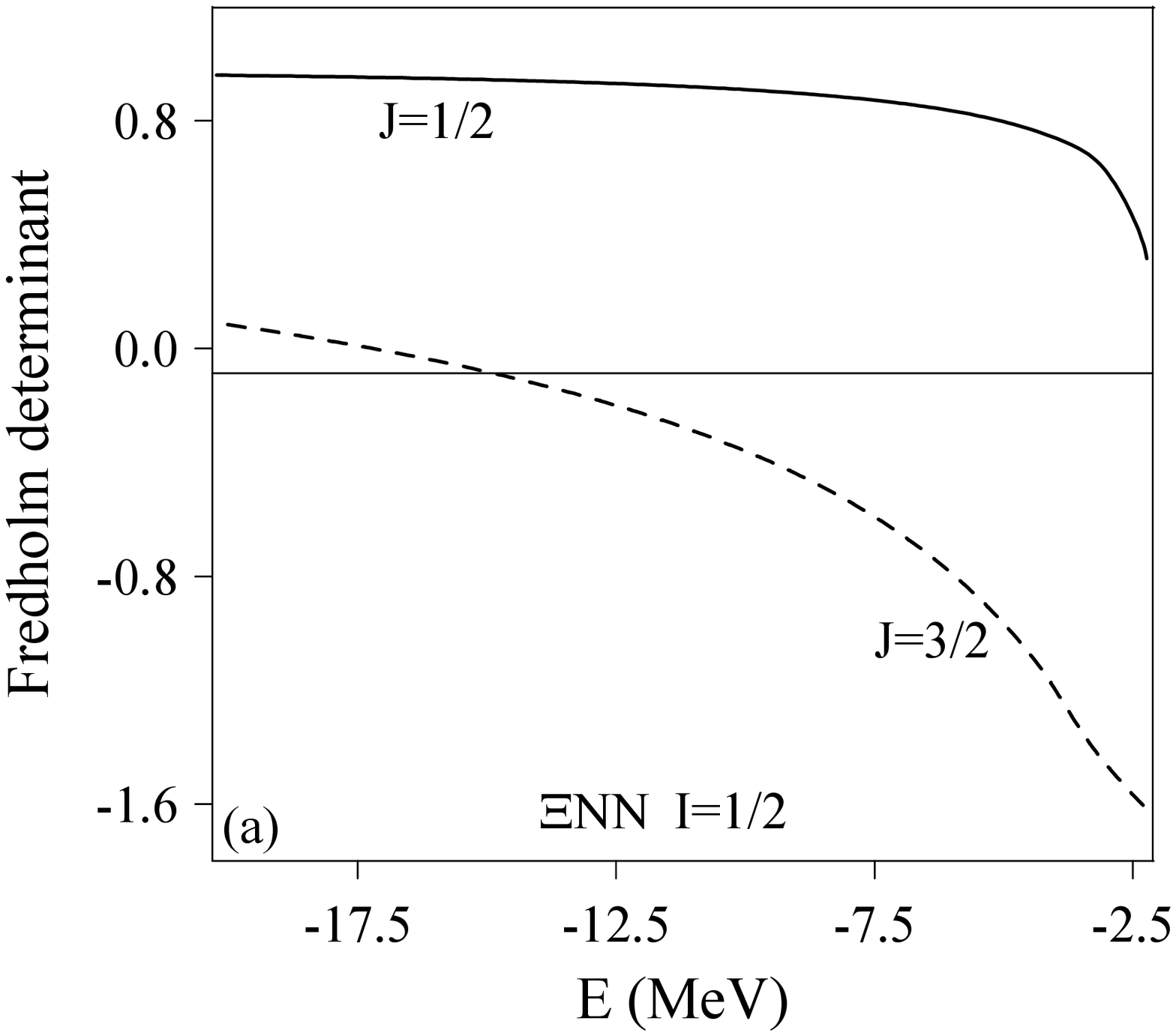}}
\resizebox{8.cm}{12.cm}{\includegraphics{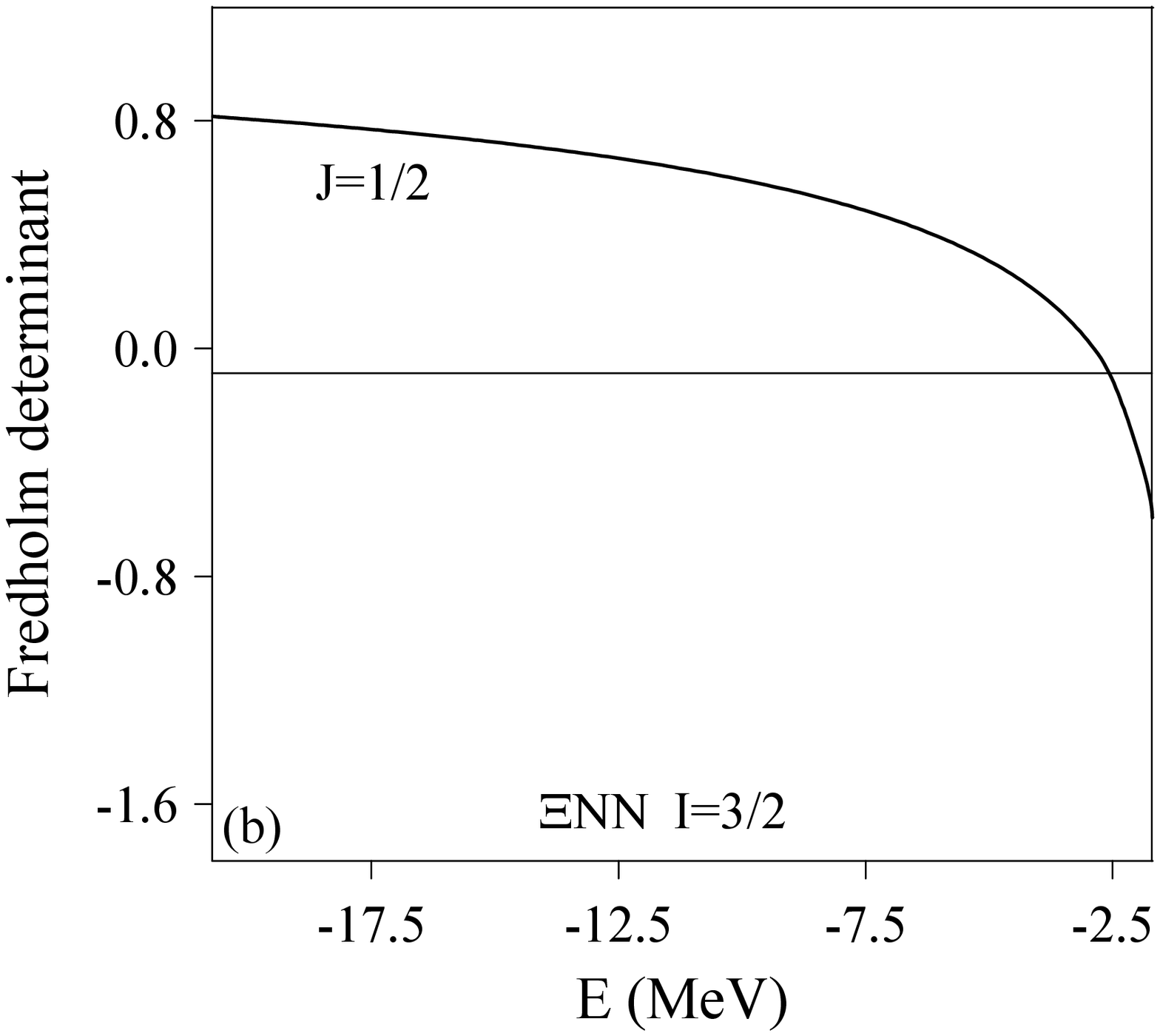}}
\vspace*{-5.0cm}
\caption{(a) Fredholm determinant for the $J=1/2$ and $J=3/2$ $I=1/2$
$\Xi NN$ channels.
(b) Fredholm determinant for the $J=1/2$ $I=3/2$
$\Xi NN$ channel.}
\label{fig2}
\end{figure*}

We show in Fig.~\ref{fig2} the Fredholm determinant
of all $\Xi NN$ channels that had been previously studied in Refs.~\cite{Gar15,Gar16}.
As we can see in Fig.~\ref{fig2}(b), a bound state is found for the
$(I)J^P=(\frac{3}{2})\frac{1}{2}^+$ $\Xi NN$ state, 1.33 MeV below
the corresponding threshold, $2m_N+m_\Xi-B_2$, where $B_2$ is the binding energy 
of the $D^*$ $\Xi N$ state. However, the most interesting result 
of the $\Xi NN$ system is shown in Fig.~\ref{fig2}(a), the very large binding
energy of the $(\frac{1}{2})\frac{3}{2}^+$ state, which would make it easy to identify 
experimentally as a sharp resonance lying some $15.7$ MeV below the $\Xi NN$ threshold.
The $\Lambda\Lambda - \Xi N$ $(i,j)=(0,0)$ transition channel, which is
responsible for the decay $\Xi NN\to\Lambda\Lambda N$, does not contribute to the 
$(I)J^P=(\frac{1}{2})\frac{3}{2}^+$ state in a pure S wave configuration~\cite{Gar16}.
One would need at least the spectator nucleon to be in a D wave or that the
$\Lambda\Lambda - \Xi N$ transition channel be in one of the 
negative parity P wave channels, with the nucleon spectator also in a P wave. 
Thus, due to the angular momentum barriers the resulting
decay width of the $(\frac{1}{2})\frac{3}{2}^+$ state is expected to be very small. 

\begin{figure*}[t]
\resizebox{8.cm}{12.cm}{\includegraphics{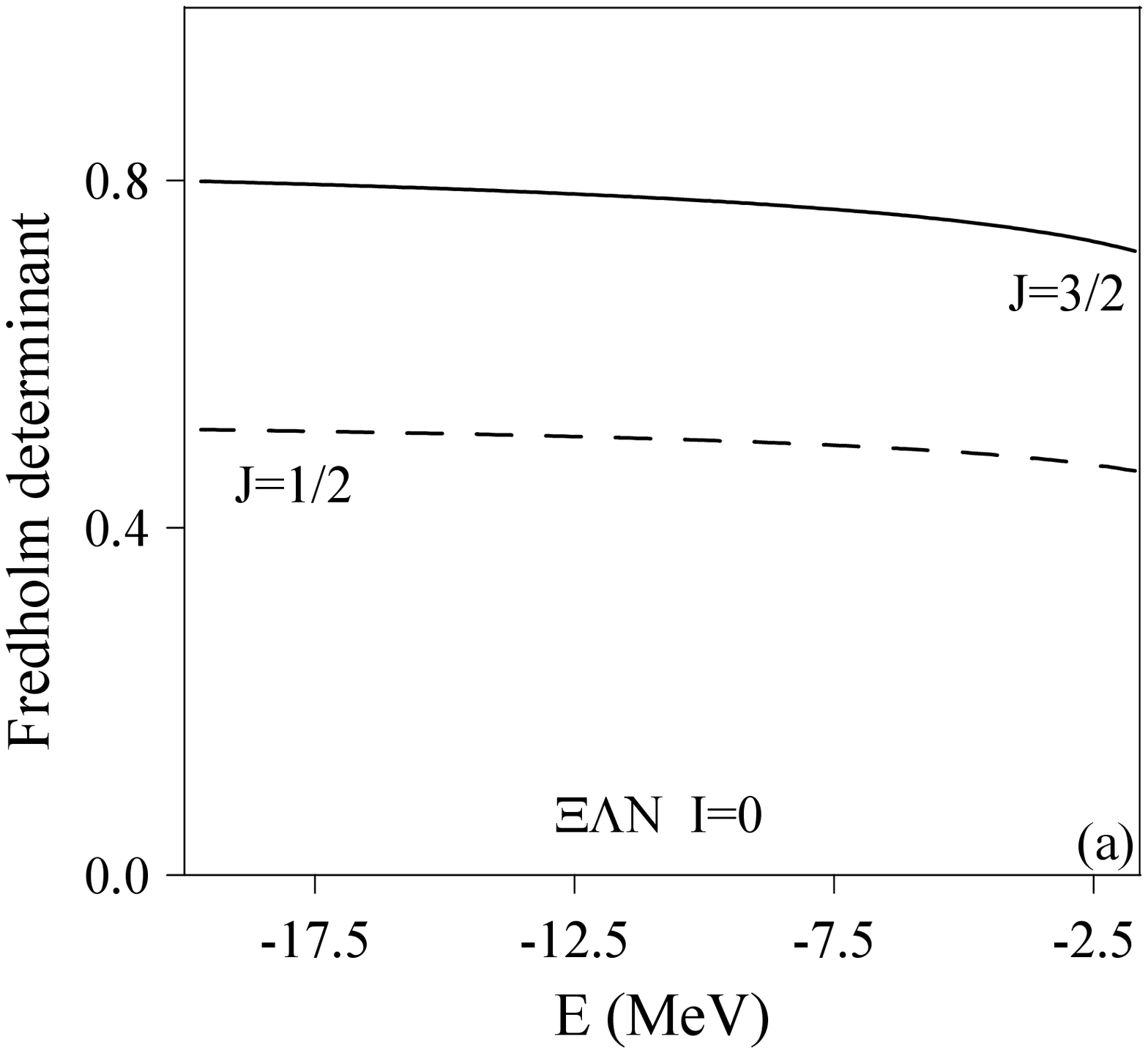}}
\resizebox{8.cm}{12.cm}{\includegraphics{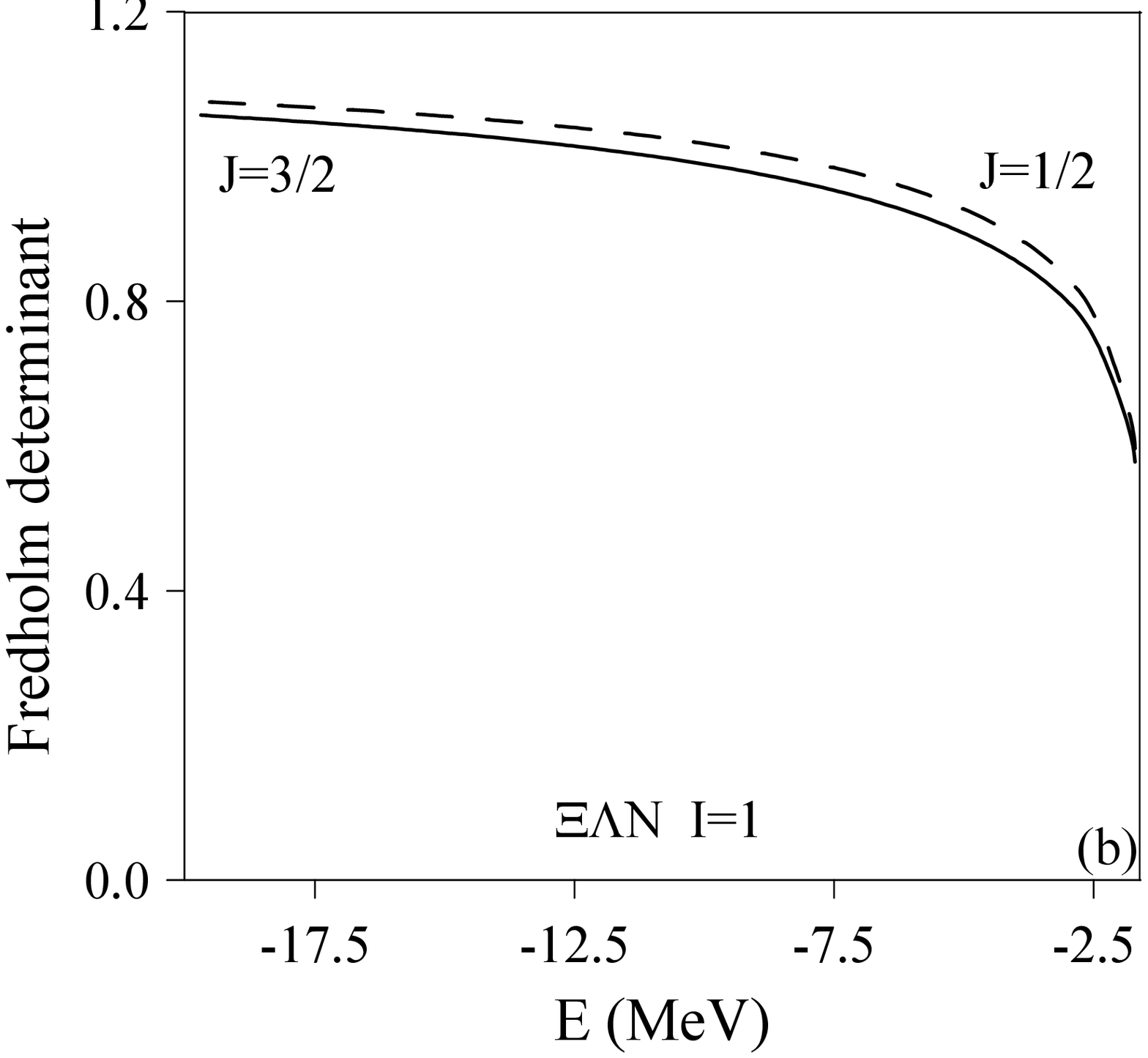}}
\vspace*{-5.0cm}
\caption{(a) Fredholm determinant for the $J=1/2$ and $J=3/2$ $I=0$
$\Xi \Lambda N$ channels.
(b) Fredholm determinant for the $J=1/2$ and $J=3/2$ $I=1$
$\Xi \Lambda N$ channels.}
\label{fig3}
\end{figure*}
\begin{figure*}[t]
\resizebox{8.cm}{12.cm}{\includegraphics{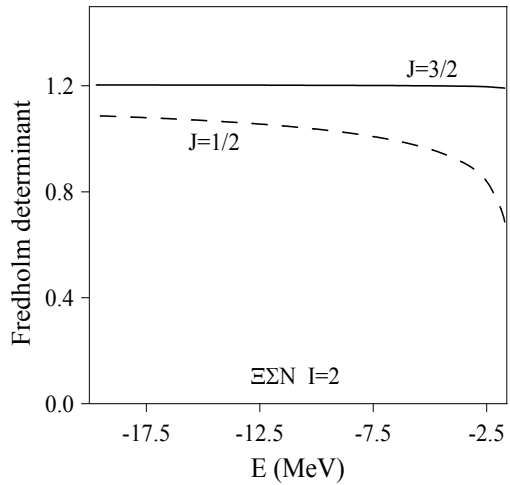}}
\vspace*{-5.0cm}
\caption{Fredholm determinant for the $J=1/2$ and $J=3/2$ $I=2$
$\Xi \Sigma N$ channels.}
\label{fig4}
\end{figure*}
\begin{figure*}[t]
\resizebox{8.cm}{12.cm}{\includegraphics{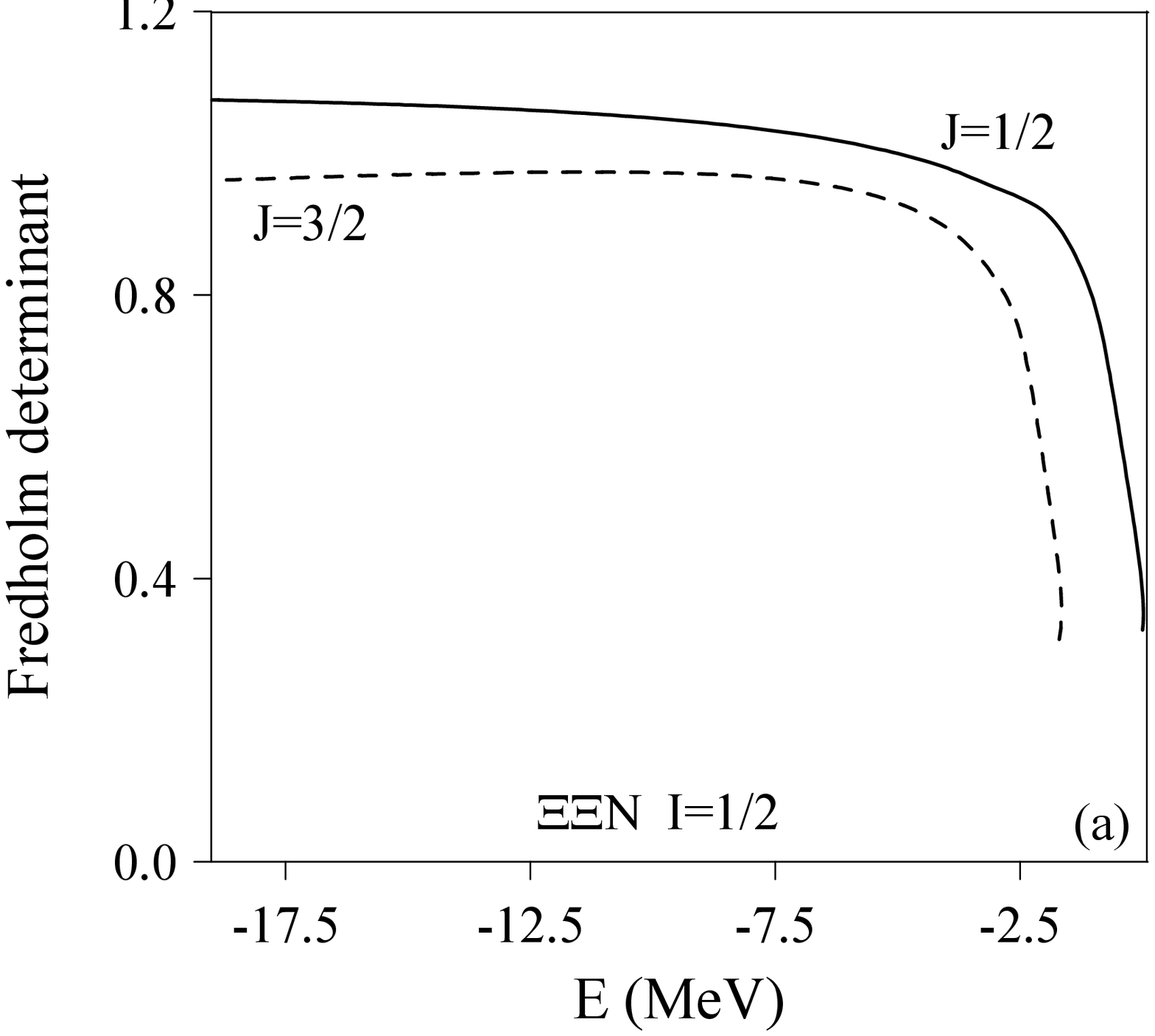}}
\resizebox{8.cm}{12.cm}{\includegraphics{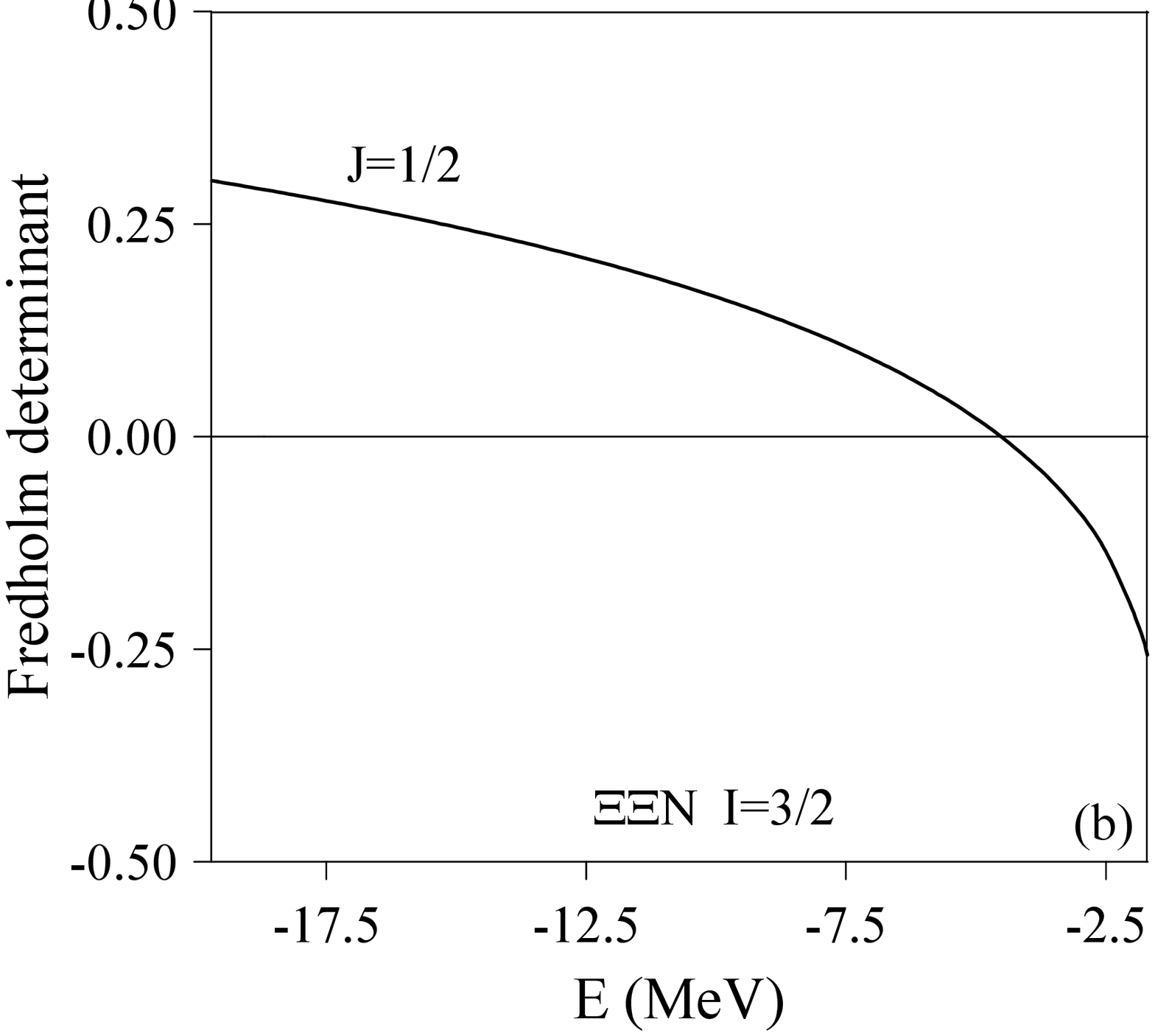}}
\vspace*{-5.0cm}
\caption{(a) Fredholm determinant for the $J=1/2$ and $J=3/2$ $I=1/2$
$\Xi\Xi N$ channels.
(b) Fredholm determinant for the $J=1/2$ $I=3/2$
$\Xi\Xi N$ channel.}
\label{fig5}
\end{figure*}

For the $\Xi NN$ three-baryon system with $(I,J)=(3/2,3/2)$, only the $(i,j)=(1,1)$ $\Xi N$ 
channel contributes (see Table~\ref{t2}),
and the corresponding Faddeev equations with two identical fermions 
can be written as~\cite{Gar07},
\begin{equation}
T=- \,t_N^{N\Xi} \, G_0 \, T \, .
\label{eqR}
\end{equation}
Thus, due to the negative sign in the r.h.s. the $\Xi N$ interaction is 
effectively repulsive and, therefore, no bound state is possible in spite of 
the attraction of the $\Xi N$ subsystem. The minus sign in 
Eq.~(\ref{eqR}) is a consequence of the identity of the two nucleons since 
the first term of the r.h.s. of Eq.~(\ref{eqR})
proceeds through $\Xi$ exchange and it corresponds to a diagram where the initial and final states 
differ only in that the two identical fermions have been interchanged which brings the minus sign. 
This effect has been pointed out before~\cite{Gar87}. This is the reason why the Fredholm 
determinant for the $(I,J)=(3/2,3/2)$ $\Xi NN$ channel is not shown in Fig.~\ref{fig2}(b).

We show in Fig.~\ref{fig3} the Fredholm determinant
of all $\Xi \Lambda N$ channels. As can be seen, although the $\Lambda N$ interaction
is attractive (see Fig.~\ref{fig1}(a)), it is not enough to generate 
bound states in the three-body system. The channels with $I=1$ are more
attractive than those with $I=0$, where the Fredholm determinant is rather flat,
but they are far from being bound. Note that whereas in the $\Xi NN$ and $\Xi\Xi N$ 
systems the $\Xi N$ interaction in the bound-state
channel appears twice, in the $\Xi\Lambda N$ system this interaction
appears only once which is the reason why this last system has
no bound states. 

We present in Fig.~\ref{fig4} the Fredholm determinant of the $I=2$
$\Xi \Sigma N$ channels. As expected, due to the contribution
of the strongly repulsive $^3S_1(I=3/2)$ $\Sigma N$ channel in all
$I=2$ $\Xi \Sigma N$ three-body systems, there do not appear any bound state.

Finally, we show in Fig.~\ref{fig5} the Fredholm determinant
of all $\Xi \Xi N$ channels. The Fredholm 
determinant for the $(I)J^P=(3/2)3/2^+$ channel is not shown in Fig.~\ref{fig5}(b)
for the same reason explained above for the $\Xi NN$ system, it is strongly repulsive.
In the $\Xi\Xi N$ system there appears a bound state with quantum numbers 
$(I)J^P=(\frac{3}{2})\frac{1}{2}^+$, 2.85 MeV below the lowest
threshold, $2m_\Xi + m_N - B_2$, where $B_2$ stands for the binding 
energy of the $D^*$ $\Xi N$ subsystem. 
Since this $\Xi\Xi N$ state has isospin $3/2$ it can not decay into
$\Xi\Lambda\Lambda$ due to isospin conservation so that it would be stable.
This stable state appears in spite of the fact
that the last update of the ESC08c Nijmegen $\Xi\Xi$ $^1S_0(I=1)$ potential 
has not bound states, as it is however predicted by several models
in the literature. If bound states would exist for the $\Xi \Xi$
system the three-body state would become deeply bound as it happens for the $\Xi NN$ system.
The $I=1/2$ channels are also attractive but they are not bound.

We summarize in Table~\ref{t3} the stable bound states of the different three-body systems
containing a $\Xi N$ subsystem.
\begin{table}[t]
\caption{Separation energy, in MeV, of the different $(I)J^P$ three-body states containing $\Xi N$ subsystems.} 
\begin{tabular}{|c|cc|} \hline\hline
$(I)J^P$ & $(\frac{1}{2})\frac{3}{2}^+$ & $(\frac{3}{2})\frac{1}{2}^+$ \\
\hline
 $\Xi NN$ & 13.54 & 1.33   \\
 $\Xi \Xi N$ &  $-$  & 2.85 \\   \hline\hline
\end{tabular}
\label{t3} 
\end{table}

\section{Summary}
\label{secV}
Recent results in the strangeness $-2$ sector, the so-called KISO event,
reported clear evidence of a deeply bound state of $\Xi^{-} - ^{14}$N
what could point out that the average $\Xi N$ interaction might be attractive. 
We have made use of the most recent updates of the ESC08c Nijmegen potential in the different
strangeness sectors, accounting for the recent experimental information, to study the 
bound state problem of three-body systems containing a $\Xi N$ subsystem: $\Xi NN$, $\Xi\Lambda N$,
$\Xi\Sigma N$ and $\Xi \Xi N$.
We have found that the $\Xi NN$ system presents bound states with quantum numbers
$(I)J^P=(3/2)1/2^+$ and $(1/2)3/2^+$, the last one being a deeply bound state lying 
13.54 MeV below the $\Xi d$ threshold. The $\Xi \Lambda N$  
system is unbound for all possible quantum numbers
due to a reduced contribution of the $\Xi N$
interaction in the bound-state channel. It occurs the same for
the $\Xi \Sigma N$ system, in this case the negative results being 
even reinforced by the contribution of the repulsive $^3S_1(I=3/2)$ $\Sigma N$ interaction.
The $\Xi\Xi N$ system presents a bound state with quantum numbers $(I)J^P=(3/2)1/2^+$. 
The states with isospin $3/2$ would be stable due to isospin conservation.
The state with isospin $1/2$ is expected to present a very small decay width
due to angular momentum barriers.
The $\Xi \Xi N$ bound state do exist in spite of the fact that we have
used the most recent update of the ESC08c Nijmegen potential that does
not predict $\Xi\Xi$ bound states. If bound states would exist for the $\Xi \Xi$
system, as predicted by several models in the literature, the state would become
deeply bound as it happens for the $\Xi NN$ system.

As stated in the introduction the hyperon-nucleon and hyperon-hyperon 
interactions are basic inputs for microscopic calculations of few- and 
many-body systems involving strangeness, such as hypernuclei or 
exotic neutron star matter. It is expected that
the recently approved hybrid experiment $E07$ at J--PARC, 
could shed light on the uncertainties of our knowledge 
of the hadron-hadron interaction in the baryon octet.
Meanwhile the scarce experimental information together with
the impossibility of microscopic calculations to study 
observations like the ones reported in Ref.~\cite{Naa15},
makes that their interpretation will be always
afflicted by large uncertainties and
gives rise to an ample room for speculation. The detailed theoretical investigation presented 
in our recent works about the possible existence of bound states 
based on realistic models are basic tools to advance in the knowledge 
of the details of the hyperon-nucleon and hyperon-hyperon interactions.
First, it could help to raise the awareness of the experimentalist 
that it is worthwhile to investigate few-baryon systems, specifically 
because for some quantum numbers such states could be stable. Secondly, 
it makes clear that strong and attractive $YN$ and $YY$ interactions,
have consequences for the few-body sector and can be easily tested
against future data.

\acknowledgments 
This work has been partially funded by COFAA-IPN (M\'exico) and 
by Ministerio de Educaci\'on y Ciencia and EU FEDER under 
Contracts No. FPA2013-47443-C2-2-P and FPA2015-69714-REDT.
A.V. is thankful for financial support from the 
Programa Propio XIII of the University of Salamanca.

\end{document}